\documentclass [aps,twocolumn] {revtex4}
\input epsf

\newcommand{\beq}{\begin{equation}}
\newcommand{\eeq}{\end{equation}}
\newcommand{\beqs}{\begin{eqnarray}}
\newcommand{\eeqs}{\end{eqnarray}}

\begin{document}

\title{Simulations of a classical spin system with competing superexchange and
double-exchange interactions}

\author {Shan-Ho Tsai\thanks{email: tsai@hal.physast.uga.edu}
and D. P. Landau\thanks{email: dlandau@hal.physast.uga.edu}}

\affiliation{
Center for Simulational Physics, University of Georgia, Athens, GA 30602}

\begin{abstract}

Monte-Carlo simulations and ground-state calculations have been used
to map out the phase diagram of a system of classical spins, on a simple 
cubic lattice, where  nearest-neighbor pairs of spins are coupled via 
competing antiferromagnetic superexchange and ferromagnetic double-exchange 
interactions. For a certain range of parameters, this model is 
relevant for some magnetic materials, such as doped manganites, which 
exhibit the remarkable colossal magnetoresistance effect.
The phase diagram includes two regions in which the two sublattice
magnetizations differ in magnitude.
Spin-dynamics simulations have been used to compute the time- and
space-displaced spin-spin correlation functions, and their Fourier
transforms, which yield the dynamic structure factor $S(q,\omega)$
for this system. Effects of the double-exchange interaction on the
dispersion curves are shown.

\end{abstract}
\maketitle


\section{Introduction}

Doped rare-earth manganites with general chemical formula Re$_{1-x}$A$_x$MnO$_3$
(where Re denotes a rare-earth and A is typically Ca or Sr)
have been shown to exhibit colossal magnetoresistance, and a wide range of other
physical properties, which are strongly dependent upon doping \cite{exp1}.
The physical mechanisms responsible for the
unusual magnetoresistance properties of these materials have not been understood
yet, and it has been suggested that double-exchange interactions \cite{zener} 
between Mn$^{+3}$ and Mn$^{+4}$ ions are important. However,
there is much ongoing debate as to whether double-exchange alone can indeed give
rise to colossal magnetoresistance or if further mechanisms have to be considered
\cite{debate}.
It is thus important to further understand properties associated with the 
double-exchange theory.

Equal valence manganese ions in the parent compounds interact via a superexchange
mechanism, whereas the different valence manganese ions, introduced through doping,
are coupled via a double-exchange interaction \cite{exp1,zener,wollan}. 
We investigate a simple model of classical spins in which nearest-neighbor spins 
interact via a competing antiferromagnetic superexchange and a ferromagnetic 
double-exchange interaction, where the ratio of these two interactions mimic
the effects of the doping $x$.
Properties of the pure Heisenberg model (only superexchange interaction) are 
well-known \cite{heis} and 
recently a high-accuracy Monte-Carlo study obtained the paramagnet-ferromagnet 
transition temperature and associated static critical exponents of the pure 
double-exchange model \cite{alvaro}. Recent studies of the phase diagram of
double-exchange systems using a different approach from ours are given in 
Ref.\cite{arovas}. We use a combination of Monte-Carlo 
simulations and ground-state calculations to determine the phase diagram of 
the model with varying ratios of the superexchange and double-exchange interaction
strengths. We also use spin-dynamics techniques to study the dynamic structure
factor and obtain the dispersion curve of the double-exchange model.

\section{Model and Methods}

The model considered here can be described by the Hamiltonian 
\beq
{\cal H}=-J_{AF}\sum_{<{\bf r},{\bf r'}>}{\bf S_r}\cdot {\bf S_{r'}}
-J_{DE}\sum_{<{\bf r},{\bf r'}>}\sqrt{1+{\bf S_r}\cdot {\bf S_{r'}}},
\label{ham}
\eeq
where ${\bf S_r}=({S_{\bf r}}^x,{S_{\bf r}}^y,{S_{\bf r}}^z)$ is a 
three-dimensional classical spin of unit length at site ${\bf r}$. The first
term, with $J_{AF}<0$, is the antiferromagnetic superexchange interaction and 
the second term describes a ferromagnetic double-exchange term with coupling
constant $J_{DE}>0$ between nearest-neighbor pairs of spins. We consider
$L\times L\times L$ simple cubic lattices with periodic boundary conditions, and
divide the system into two sublattices, $A$ and $B$, with magnetizations ${\bf m}_A$ 
and ${\bf m}_B$, respectively. The scalar product is defined as
${\bf m}_A \cdot {\bf m}_B= |{\bf m}_A| |{\bf m}_B| \cos{\theta}\equiv y$, where clearly
$-1\leq y \leq 1$. 

In the ground state, the condition of energy minimization 
leads to the expression
\beq
y=(J_{DE})^2/(4|J_{AF}|^2)-1,
\label{yeq}
\eeq
with $|J_{AF}|/J_{DE}\geq 1/\sqrt{8}=0.35355...$, 
and for $|J_{AF}|/J_{DE} < 0.35355...$ the ground state is ferromagnetic.
Eq.(\ref{yeq}) admits the
possibility of non-colinear phases and/or sublattice magnetizations with different
magnitudes. Note that $y=0$ for $|J_{AF}|/J_{DE}=1/2$, which means that $\theta=\pi/2$
and/or that the magnetization in at least one sublattice is zero in this case.

For $T>0$ we use Monte-Carlo simulations to measure thermodynamic quantities of
the model in order to determine the phase diagram. We used the Metropolis algorithm, 
lattice sizes $L=12$ and $24$, and typically we discarded 10000 Monte-Carlo steps for
thermalization and used 10000-30000 steps in computing averages.

The dynamics of the spins are governed by the equations of motion
\beq
\frac{d}{dt}{\bf S_r}={\bf S_r} \times (-\vec{\bigtriangledown}_{\bf r}{\cal H}),
\label{eqmotion}
\eeq
where the effective field
\beq
-\vec{\bigtriangledown}_{\bf r}{\cal H}=J_{AF}\sum_{\bf r'}{\bf S_{r'}}+\frac{J_{DE}}{2}
\sum_{\bf r'}\frac{\bf S_{r'}}{\sqrt{1+{\bf S_r}\cdot {\bf S_{r'}}}}
\eeq
and the sums are over ${\bf r'}$ nearest-neighbor to ${\bf r}$. These equations were 
solved numerically using a method based on second-order Suzuki-Trotter decomposition
of exponential operators \cite{krech}, to a maximum time of $t_{max}=440/J_{DE}$. 

The dynamic structure factor $S({\bf q},\omega)$ for momentum transfer
${\bf q}$ and frequency transfer $\omega$, observable in neutron scattering
experiments, is given by
\beq
S^k({\bf q},\omega)=\sum_{{\bf r,r'}} \exp[i {\bf q}\cdot ({\bf r}-{\bf r'})]
\int_{-\infty}^{+\infty} \exp(i\omega t) C^k({\bf r} - {\bf r'},t) \frac{dt}{\sqrt{2\pi}},
\eeq
where $C^k({\bf r} - {\bf r'},t)$ is the space-displaced, time-displaced spin-spin correlation
function defined, with $k=x, y,$ or $z$, as
$
C^k({\bf r} - {\bf r'},t) =\langle {S_{{\bf r}}}^k(t){S_{{\bf r'}}}^k(0)\rangle-
\langle {S_{{\bf r}}}^k(t)\rangle\langle {S_{{\bf r'}}}^k(0)\rangle.
$
The displacement ${\bf r}$ is in units of the lattice unit cell length $a$. More
details on the spin-dynamics methods are given in Refs. \cite{kun,alex}.

\section{Results}

Our preliminary results show that $|{\bf m}_A|=|{\bf m}_B|$ at all temperatures for
$|J_{AF}|/J_{DE}<1/\sqrt{8}$, decreasing smoothly
from $|{\bf m}_A|=|{\bf m}_B|=1$ at low $T$, towards zero as $T$ increases. For 
$|J_{AF}|/J_{DE}>1/\sqrt{8}$ we find that at low temperatures $|{\bf m}_A|\ne |{\bf m}_B|$,
as illustrated in Fig. 1 for $|J_{AF}|/J_{DE}=0.9$.
For this range of interaction strengths 
our results indicate that $|{\bf m}_A|=|(J_{DE})^2/(4|J_{AF}|^2)-1|$ and $|{\bf m}_B|=1$ at
$T=0$, and Eq.(\ref{yeq}) then implies that $\theta =0$ for $|J_{AF}|/J_{DE}<1/2$, 
$\theta =\pi$ for $|J_{AF}|/J_{DE}>1/2$, and the sublattice
magnetizations reduce to $|{\bf m}_A|=0$ and $|{\bf m}_B|=1$ for $|J_{AF}|/J_{DE}=1/2$. 
The phase-transition temperatures are obtained from analyses of the sublattice
magnetizations for $L=24$, hence there are some uncertainties in these 
estimates due to finite-size effects. 
\begin{figure}
\centering
\leavevmode
\epsfxsize=8.0cm
\epsffile{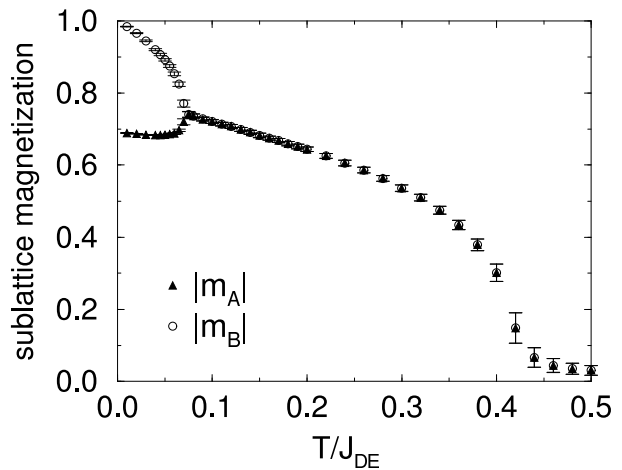}
\caption{Sublattice magnetization for $|J_{AF}|/J_{DE}=0.9$, and lattice size $L=24$.}
\label{mags0p9}
\end{figure}

The phase diagram that we obtained from the analysis of the sublattice magnetizations, shown in
Fig. 2, has five different regions: (i) ferromagnetic (F) for low $T$ and
small values of $|J_{AF}|/J_{DE}$; (ii) 
antiferromagnetic (AF) for moderate $T$ and high values of $|J_{AF}|/J_{DE}$;
(iii) paramagnetic (PM) at high $T$; and two low
temperature phases, which we label as (iv) region I for $1/\sqrt{8}<|J_{AF}|/J_{DE}<1/2$,
and (v) region II for $|J_{AF}|/J_{DE}>1/2$.
Regions (i)-(iii) are characterized by $|{\bf m}_A|=|{\bf m}_B|$, with ${\bf m}_A={\bf m}_B$ 
in the ferromagnetic
phase and ${\bf m}_A=-{\bf m}_B$ in the antiferromagnetic phase. In contrast, in regions
I and II we find that $|{\bf m}_A|\ne |{\bf m}_B|$ and although our simulations suggest that
$\theta =0$ and $\pi$ in regions I and II, respectively, we cannot rule out possible small
deviations from the aligned and anti-aligned sublattice spin configurations in the respective
regions. 
The question marks (?) in the phase diagram (see Fig. 2) indicate that the
points at which the F-PM and AF-PM transition lines join other phase boundaries have not
been determined accurately yet. 
For the pure double-exchange ferromagnetic model ($J_{AF}=0$), the critical temperature
(F-PM transition point) is ${T_c}^{DE}=0.74515(7)J_{DE}$ \cite{alvaro}, whereas
for the pure Heisenberg model $T_c^{SE}=1.442929(77)|J_{AF}|$ \cite{heis}
for either ferromagnetic or antiferromagnetic interaction. 
\begin{figure}
\centering
\leavevmode
\epsfxsize=8.0cm
\epsffile{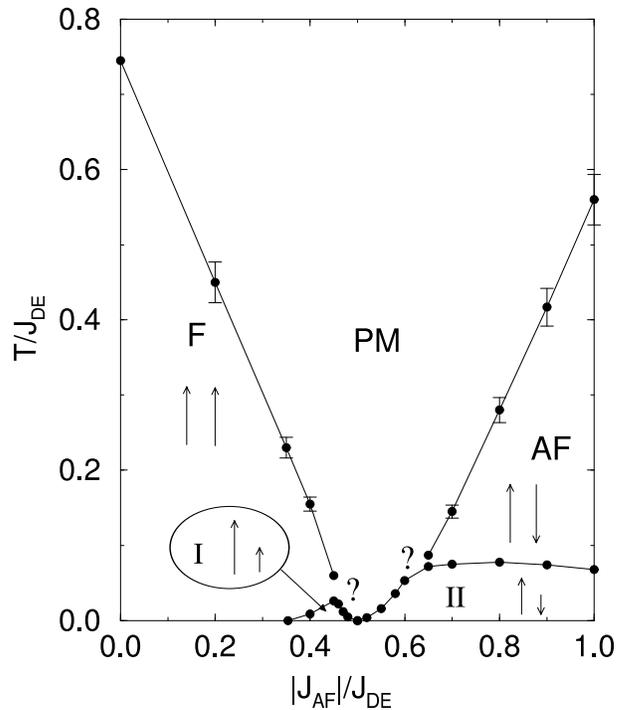}
\caption{Phase diagram. When not shown, the error bars are smaller than or of the size of the points and the solid lines simply join the data, guiding the eyes.}
\label{phasediag}
\end{figure}

Fig. 3 shows a comparison of the dispersion relations for the pure
Heisenberg ferromagnet and for the pure double-exchange ferromagnetic model.
The results at $T=0$ correspond to the linear spin-wave theory, and data at the
respective critical temperatures are from our simulations. 
\begin{figure}
\centering
\leavevmode
\epsfxsize=8.0cm
\epsffile{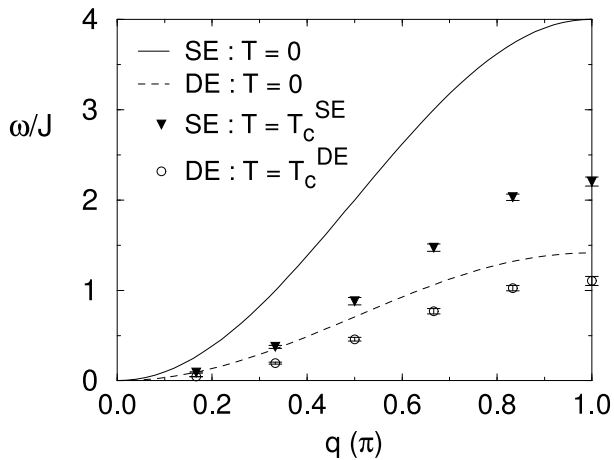}
\caption{Dispersion relations for the Heisenberg ferromagnet (pure superexchange model, labeled as SE) and the pure ferromagnetic double-exchange model (DE).}
\label{disprel}
\end{figure}

\section{Conclusions}

We studied a system of classical spins coupled via competing antiferromagnetic
superexchange and ferromagnetic double-exchange interactions. As expected, when the
former is much weaker than the latter, the system is ferromagnetic and in the 
reverse case the system is antiferromagnetic. In addition, we found two other low
temperature phases, in which the sublattice magnetizations differ in magnitude and
are close to being aligned or anti-aligned, depending on the ratio of the two
interaction strenghts. At high temperature the system is paramagnetic. 
Comparing the dispersion curves we see that spin-waves in the pure double-exchange
model occur at lower frequency transfer than in the pure Heisenberg ferromagnet.

\begin{center}
{\bf Acknowledgments}
\end{center}
This research was supported in part by NSF grant No. DMR-9727714. 


\end{document}